# A new multiscale modeling approach to unravel the influence of interlayer sp$^3$ bonds on the nonlinear large-deformation and fracture behaviors of 2D carbon nanostructures under tension

Xiangyang Wang[*], Huibo Qi, Biao Xu, Shichao Dai, Jiqiang Li

*School of Transportation, Ludong University, Yantai 264025, China*

**Abstract**

To delve deeply into the nonlinear large-deformation and fracture behaviors of 2D carbon nanostructures (2D CNs), including bilayer graphene, diamane, and their transitional structures, this paper introduces a multiscale auxiliary nodes (MAN) method rooted in atomic structures and potentials. This approach simulates 2D CNs by constructing two virtual continuum sheets with high-order continuity. The moving least squares (MLS) approximation is employed to facilitate the transformation between atomic displacements and nodal displacements, thereby converting atomic potential energy into strain energy within the continuum model. Through iterative solutions of nonlinear stiffness equations, the equilibrium configuration of the system under specified loading conditions can be obtained. The flexibility in the density and arrangement of nodes allows for a smooth and seamless cross-scale transition from discrete atomic structures to a continuum model. Numerical simulations demonstrate that MAN method accurately predicts the nonlinear large-deformation and fracture behaviors of 2D CNs. The Young's modulus and shear modulus of diamane in both zigzag and armchair directions closely approach those of diamond and are notably higher than those of graphene. Furthermore, the quantity and distribution of interlayer sp³ bonds significantly influence the fracture behavior of 2D CNs, with strategic placement of these bonds effectively enhancing the tensile strength of the structures.

[*] Corresponding author.
*E-mail address:* wxy017@ldu.edu.cn (X. Wang).

## 1. Introduction

Since its first successful isolation via mechanical exfoliation in 2004, graphene has attracted extensive attention within the scientific community.[1] Graphene, a single layer of graphite bonded through sp² hybridized orbitals, exhibits a series of remarkable physical and mechanical properties, including high thermal conductivity, unique electrical characteristics, and exceptional mechanical strength.[2-4] Diamane, a 2D nanostructure that shares similarities with bilayer graphene, was first theoretically predicted in 2009. Compared with bilayer graphene, the most distinctive feature of diamane is the presence of full interlayer sp³ covalent bonds that connect all the carbon atoms within the structure. Through employing high-pressure techniques, the bonding state of carbon atoms can be altered from sp² hybridization to sp³ hybridization, thus enabling the conversion of bilayer graphene into pristine diamane.[5-9] As a 2D carbon nanostructure (CN), diamane demonstrates superior physical and mechanical performance.[10,11] Specifically, the Young's modulus and shear modulus of pristine diamane are consistent with those of diamond and exceed the corresponding values of graphene, being nearly twice as high as those of hydrogenated diamane.[12] Furthermore, AA-stacked diamane and AB-stacked diamane exhibit nearly identical Young's modulus, shear modulus, Poisson's ratio, and fracture strength.[12,13]

In the academic realm, a substantial number of theoretical and computational studies have been carried out with the aim of predicting the properties of bilayer or few-layer graphene with interlayer sp³ bonds. Through laser annealing treatment,[14] few-layer graphene film samples with chemical modification characteristics have been successfully fabricated. This material is characterized by randomly distributed sp³ bonds connecting the graphene layers and exhibits excellent mechanical strength. Research has demonstrated that by inserting sp³ bonds in an ordered manner to form superlattices in bilayer graphene with specific symmetry and periodicity, its electronic properties can be precisely regulated.[15,16] Moreover, through the deliberate adjustment of the density and spatial distribution of interlayer sp³ bonds, the thermal conductivity[17,18] and resonance characteristics[19,20] of the resultant materials can be effectively manipulated. The mechanical properties of bilayer graphene/diamane can also be modulated by regulating the density and distribution of interlayer sp³ bonds.[21] Studies indicate that sp³ bonds play a pivotal role in augmenting the interlayer shear modulus and load transfer rate of bilayer graphene, while simultaneously enhancing its stability when subjected to axial compression.[22] In addition to strengthening the mechanical properties of bilayer graphene, interlayer sp³ bonds can also enhance the load transfer capacity of double-walled carbon nanotubes. This, in turn, significantly improves their anti-buckling and anti-bending performance while reducing their

sensitivity to defects.[23-26]

Currently, the research methods employed for investigating 2D carbon nanostructures (2D CNs), such as bilayer graphene, diamane, and their intermediate transition structures (where only a portion of the carbon atoms form interlayer $sp^3$ hybridization), predominantly rely on atomic computational approaches, including molecular dynamics simulations and density functional theory.[14-29] In contrast, there are relatively few developed multiscale methods available for studying the mechanical properties of 2D CNs.[30] Compared with atomic methods, multiscale methods break through the limitations of atomic methods in terms of scale, time, and resource consumption while maintaining critical physical accuracy. Consequently, they have emerged as a core tool for studying complex systems (e.g., macromolecules, micro/nano-scale structures, and functional materials) and long-timescale phenomena (e.g., phase transitions and self-assembly).[31-33] The advantages of multiscale methods are not merely reflected in computational efficiency; more importantly, they provide a systematic framework for understanding cross-scale mechanisms and enabling rational design.

This paper aims to develop a multiscale auxiliary nodes (MAN) method based on atomic structures and atomic potentials to study the nonlinear large-deformation and fracture behaviors of 2D CNs. In this method, 2D CNs, including bilayer graphene, diamane, and their intermediate transition structures, are simulated using two virtual continuum sheets with higher-order continuity. The fundamental representative cell of the 2D CNs is selected to calculate the atomic potential energy, which is then mapped to the strain energy in the continuum method. Nodes are placed on the virtual continuum sheets, and influence domains are defined for each atom. A displacement interpolation relationship is established between the nodes and atomic points, thereby transforming the total potential energy functional expressed in terms of atomic displacements into one expressed in terms of nodal displacements. Given the relatively free placement of nodes, the number of degrees of freedom of the studied system can be freely adjusted. This enables the achievement of improved computational efficiency while ensuring computational accuracy.

## 2. Multiscale auxiliary nodes (MAN) method

In this section, we present a comprehensive elaboration on the construction process of MAN method. Firstly, two parallel virtual continuum sheets with higher-order continuity are constructed. The atomic structure and interatomic interactions of 2D CNs are mapped onto these continuum sheets. Subsequently, a potential energy functional with nodal displacements as independent variables is established. Finally, the equilibrium configuration of the system under given loading conditions is obtained through iterative solution of the nonlinear stiffness equations.

### 2.1 Multiscale model

From a configurational perspective, bilayer graphene, diamane, and their intermediate transition structures exhibit remarkable similarity, namely, all are composed of two layers of carbon atoms. The primary distinction lies in whether the carbon atoms are interconnected through $sp^2$ or $sp^3$ hybridization. Building upon the structural characteristics of these three materials, we propose a hypothetical "bilayer graphene + interlayer bonding" model, as illustrated in Fig. 1(a). The sp³ hybridized configuration of carbon atoms is depicted in Fig. 1(b), whereas the sp² hybridized structure can be regarded as a degenerate form derived from the sp³ hybridized structure by removing certain C-C bonds, as shown in Fig. 1(a). Consequently, this paper uniformly adopts the sp³ hybridized structure shown in Fig. 1(b) as the representative cell for the aforementioned three materials, and in cases involving sp² hybridization, it will be modeled by removing specific C-C bonds.

Assuming that carbon atom displacements can be described by a higher-order continuous displacement field, two virtual continuum sheets with higher-order continuity are constructed, as depicted in Fig. 1(a). The higher-order continuity of these sheets ensures a smooth transition between different regions of the nanostructure, which is crucial for accurately simulating large-deformation behavior. Given the problem characteristics, nodes can be freely arranged on the two continuum sheets, providing flexibility in adjusting the system's degrees of freedom. Influence domains are then assigned to each atom, as illustrated in Fig. 1(c). Atomic displacements can be interpolated from the nodes within their respective influence domains. The moving least squares (MLS) approximation[34] which is widely used in meshless methods[35–37] for the construction of interpolation function with higher-order continuity, will be employed for cross-scale displacement interpolation. Consequently, this modeling approach allows for the efficient representation of complex atomic arrangements and the incorporation of atomic-scale interactions into the continuum-scale framework.

The distance between two bonded carbon atoms *I* and *J* can be described by

$$r_{IJ} = |\boldsymbol{x}_I - \boldsymbol{x}_J|, \tag{1}$$

where $\boldsymbol{x}_I$ and $\boldsymbol{x}_J$ represent the positions of the carbon atoms *I* and *J* in the current configuration, respectively. These positions can be further expressed as

$$\boldsymbol{x}_I = \boldsymbol{X}_I + \boldsymbol{u}_I, \tag{2}$$

$$\boldsymbol{x}_J = \boldsymbol{X}_J + \boldsymbol{u}_J, \tag{3}$$

where $\boldsymbol{X}_I$ and $\boldsymbol{X}_J$ denote the coordinates of carbon atoms *I* and *J* in the initial configuration (graphene sheet), respectively. $\boldsymbol{u}_I$ and $\boldsymbol{u}_J$ are the displacements of atoms *I* and *J*, respectively, from the initial configuration to the current configuration.

The internal energy of the 2D atomic system can be written as

$$U_{\text{int}} = \sum_{i=1}^{N} V_0(r_{IJ}), \tag{4}$$

where *N* denotes the total number of cells in the atomic system and $V_0$ represents the atomic potential energy per cell (see Fig. 1(b)), which can be described by the second-generation reactive empirical bond order (REBO) potential.[38] The REBO potential is a widely used empirical potential capable of accurately modeling covalent bonding and angular dependence in carbon-based materials. Once the atomic potential energy is calculated, it is mapped to the strain energy in the continuum model. This mapping is achieved by establishing a relationship between atomic displacements and the strain field in the continuum sheets. The strain energy in these sheets is then expressed as a function of the nodal displacements using the moving least squares (MLS) approximation as the interpolation function, where nodal displacements are the primary variables in the multiscale method. The undeformed equilibrium configuration of the 2D CNs under an unloaded state is obtained by minimizing the potential energy functional with respect to nodal displacements (see Fig. 2 for an illustration).

**2.2 MLS approximation**

MLS approximation is a powerful numerical technique used for fitting a smooth, approximate function to a set of scattered data points.[34] In MLS approximation, a variable of interest (e.g., a displacement field function) is approximated by a weighted sum of polynomial basis functions. For a 2D system, the approximation can be expressed as

$$u^h(X) = \sum_{i=0}^{m} p_i(X)\, a_i(X) = \boldsymbol{p}^{\mathrm{T}}(X)\boldsymbol{a}(X), \tag{5}$$

where $\boldsymbol{p}(X)$ represents the complete polynomial basis function of order $m$, and $\boldsymbol{a}(X)$ are the non-constant coefficients, which are functions of $X$ and vary spatially.

The unknown coefficients $\boldsymbol{a}(X)$ can be determined by minimizing of the weighted discrete $L_2$ norm, which measures the difference between the approximate function and the actual data points, as

$$J(\boldsymbol{a}) = \sum_{I=1}^{N} w_I(X)\, (\boldsymbol{p}^{\mathrm{T}}(X_I)\boldsymbol{a}(X) - \tilde{u}_I)^2, \tag{6}$$

where $N$ is the number of nodes in the influence domain. $w_I(X)$ represents the weight function associated with node $I$, which is non-zero only within its compact support. $\tilde{u}_I$ denotes the nodal parameter at node $I$.

Minimizing $J(\boldsymbol{a})$ with respect to $\boldsymbol{a}$ leads to a set of linear equations as

$$\boldsymbol{A}(\boldsymbol{X})\boldsymbol{a}(\boldsymbol{X}) = \boldsymbol{C}(\boldsymbol{X})\widetilde{\boldsymbol{u}}, \tag{7}$$

where $\widetilde{\boldsymbol{u}} = (\tilde{u}_1, \tilde{u}_2, \ldots, \tilde{u}_N)^{\mathrm{T}}$,

$$\boldsymbol{A}(\boldsymbol{X}) = \sum_{I=1}^{N} w_I(\boldsymbol{X})\boldsymbol{p}(\boldsymbol{X}_I)\boldsymbol{p}^{\mathrm{T}}(\boldsymbol{X}_I), \tag{8}$$

and

$$\boldsymbol{C}(\boldsymbol{X}) = [w_1(\boldsymbol{X})\boldsymbol{p}(\boldsymbol{X}_1), w_2(\boldsymbol{X})\boldsymbol{p}(\boldsymbol{X}_2), \cdots, w_N(\boldsymbol{X})\boldsymbol{p}(\boldsymbol{X}_N)]. \tag{9}$$

The coefficients $a(\boldsymbol{X})$ can be calculated by solving Eq. (7) as

$$\boldsymbol{a}(\boldsymbol{X}) = \boldsymbol{A}^{-1}(\boldsymbol{X})\boldsymbol{C}(\boldsymbol{X})\widetilde{\boldsymbol{u}}. \tag{10}$$

Substituting Eq. (10) into Eq. (5) yields the standard form of the MLS approximation as

$$u^h(\boldsymbol{X}) = \sum_{I=1}^{N} \phi_I(\boldsymbol{X})\tilde{u}_I = \boldsymbol{\phi}^{\mathrm{T}}(\boldsymbol{X})\widetilde{\boldsymbol{u}}, \tag{11}$$

where $\phi_I(\boldsymbol{X})$ is the MLS shape function for node $I$, calculated as

$$\phi_I(\boldsymbol{X}) = \boldsymbol{p}^{\mathrm{T}}(\boldsymbol{X})\boldsymbol{A}^{-1}(\boldsymbol{X})\boldsymbol{C}(\boldsymbol{X})_I. \tag{12}$$

The weight function $w_I(\boldsymbol{X})$ plays a crucial role in MLS approximation, as it determines the contribution of each node to the approximation at a given point $\boldsymbol{X}$. The weight function is typically chosen to have compact support, meaning it is non-zero only within a small neighborhood of node $I$. In this study, a cubic spline is used as the weight function.

The size of the influence domain of node $I$ can be defined as

$$D_I = s \cdot \max_{J \in N_I} \|\boldsymbol{X}_I - \boldsymbol{X}_J\|, \tag{13}$$

where $N_I$ is the set of neighboring nodes of node $I$. $s$ is a scaling parameter generally greater than 2.0, ensuring the matrix $A(X)$ is invertible.

**2.3 REBO potential for carbon**

The REBO potential, as proposed by Brenner et al.,[38] serves as a widely adopted framework for characterizing the interactions between carbon atoms within CN. This potential adopts the following mathematical formulation as

$$V(r_{IJ}) = V_R(r_{IJ}) - \bar{B}_{IJ} V_A(r_{IJ}), \quad (14)$$

where $r_{IJ}$ is the distance between atoms $I$ and $J$, and the repulsive ($V_R$) and attractive ($V_A$) terms are specified as

$$V_R(r_{IJ}) = f^c(r_{IJ}) \left(1 + \frac{Q}{r_{IJ}}\right) A \exp(-\alpha r_{IJ}), \quad (15)$$

$$V_A(r_{IJ}) = f^c(r_{IJ}) \sum_{n=1,3} B_n \exp(-\beta_n r_{IJ}). \quad (16)$$

The parameters $Q$, $A$, $\alpha$, $B_n$, $\beta_n$ are obtained through data-fitting procedures.[38] The function $f^c(r_{IJ})$ denotes a smooth cutoff function designed to confine the range of covalent interactions, typically expressed as

$$f^c(r_{IJ}) = \begin{cases} 1 & r_{IJ} < R_{\min} \\ \frac{1}{2}\left\{1 + \cos\left(\frac{r_{IJ} - R_{\min}}{R_{\max} - R_{\min}}\right)\right\} & R_{\min} < r_{IJ} < R_{\max}, \\ 0 & r_{IJ} > R_{\max} \end{cases} \quad (17)$$

where $R_{\min} = 1.7$ Å, and $R_{\max} = 2.0$ Å. It is crucial to acknowledge that while the cutoff function $f^c(r_{IJ})$ effectively restricts the range of interactions, it may introduce artifacts such as nonphysical increases in bond force within certain distance ranges (e.g., 1.7 to 2.0 Å).[39,40] To mitigate this, the application of the cutoff function is limited to initial and undeformed configurations where the C-C bond lengths remain below a predefined threshold, ensuring the physical fidelity of the simulation results.

The empirical bond order function, $\bar{B}_{IJ}$, which modulates the strength of the bond based on its local environment, is computed as

$$\bar{B}_{IJ} = \frac{1}{2}\left(B_{IJ}^{\sigma-\pi} + B_{JI}^{\sigma-\pi}\right) + \pi_{IJ}^{RC} + B_{IJ}^{DH}. \tag{18}$$

Herein, $\pi_{IJ}^{RC}$ accounts for the influence of radical energetics and $\pi$-bond conjugation effects on the bond energies, and $B_{IJ}^{DH}$ relies on the torsion angle for the C-C bonds of non-conjugated double bond system. For bonds outside the conjugated system and planar double bonds, both terms are zero. $B_{IJ}^{\sigma-\pi}$ depends on the bond angles and local coordination of atoms $I$ and $J$, calculated by

$$B_{IJ}^{\sigma-\pi} = \left\{1 + \sum_{K \neq I,J} f_{IK}^c(r_{IK}) G(\cos(\theta_{IJK}))\right\}^{-\frac{1}{2}}, \tag{19}$$

where $\theta_{IJK}$ represents the bond angle between atoms $I$, $J$, and $K$, and $G(\cos(\theta_{IJK}))$ is a sixth-order polynomial spline interpolation function that adjusts the contribution of each nearest neighbor to the bond order according to the cosine of the bond angle.

**2.4 Numerical computational scheme for MAN model**

As mentioned above, the sp³ hybridized structure was selected as the representative cell for the 2D nanostructures (see Fig. 1(b)). To ensure its compatibility with both sp² and sp³ hybridization simultaneously, the REBO potential for the representative cell can be written as

$$V_0(r) = e_1 V_R(r_1)$$

$$-\frac{1}{2}\left(\left\{1 + \sum_{i=2|_1}^{4|_1} e_i f^c(r_i) G(\cos(\theta_{1i}))\right\}^{-\frac{1}{2}}\right.$$

$$\left.+ \left\{1 + \sum_{i=5|_1}^{7|_1} e_i f^c(r_i) G(\cos(\theta_{1i}))\right\}^{-\frac{1}{2}}\right) V_A(r_1), \tag{20}$$

where the coefficients $e_1$ to $e_7$ take values of 0 or 1, They are used to control the inclusion or exclusion of bonds and bond angles, thereby enabling the adjustment between sp² and sp³ hybridization. For instance, when all $e$ values are set to 1, it corresponds to sp³ hybridization. The notation $j|_i$ denotes the $j$-th bond that is connected to the bond $i$ within the same cell. $\theta_{1i}$ represents the angle between the bonds $r_1$ and $r_i$.

By combining Eq. (4) and Eq. (20), the internal energy of 2D nanostructures can be rewritten as

$$U_{\text{int}} = \sum_{i=1}^{N} e_i V_R(r_i)$$

$$-\frac{1}{2}\left(\left\{1 + \sum_{j=2|_i}^{4|_i} e_i f^c(r_i) G(\cos(\theta_{1i}))\right\}^{-\frac{1}{2}}\right.$$

$$\left.+ \left\{1 + \sum_{j=5|_i}^{7|_i} e_i f^c(r_i) G(\cos(\theta_{1i}))\right\}^{-\frac{1}{2}}\right) V_A(r_i). \tag{21}$$

This study focuses on the tensile mechanical properties of 2D nanostructures, neglecting the interlayer van der Waals interactions. Consequently, the total potential

energy functional of the atomic system can be expressed as

$$U_{\text{tot}}(\boldsymbol{u}) = U_{\text{int}}(\boldsymbol{u}) - \sum_{I=1}^{M} \boldsymbol{F}_I \cdot \boldsymbol{u}_I, \tag{22}$$

where $\boldsymbol{F}_I$ represents the external force (if applicable) acting upon atom $I$.

Taking into account the established MAN model, which connects the atomic-scale and continuum-scale descriptions of the nanostructures, with MLS as the transformation mapping, $U_{\text{tot}}$ can be described as a function of the nodal displacement parameters $\widetilde{\boldsymbol{u}}$ as

$$U_{\text{tot}}(\widetilde{\boldsymbol{u}}) = U_{\text{int}}(\widetilde{\boldsymbol{u}}) - \sum_{I=1}^{M} \boldsymbol{F}_I \cdot \boldsymbol{u}_I^h(\widetilde{\boldsymbol{u}}). \tag{23}$$

The equilibrium solution can be obtained through energy minimization as

$$\frac{\partial U_{\text{tot}}(\widetilde{\boldsymbol{u}})}{\partial \widetilde{\boldsymbol{u}}} = \boldsymbol{0}. \tag{24}$$

Due to the nonlinear nature of Eq. (24), a Taylor expansion is performed on it, as

$$\left.\frac{\partial U_{\text{tot}}(\widetilde{\boldsymbol{u}})}{\partial \widetilde{\boldsymbol{u}}}\right|_{\widetilde{\boldsymbol{u}}=\widetilde{\boldsymbol{u}}^n} + \left.\frac{\partial^2 U_{\text{tot}}(\widetilde{\boldsymbol{u}})}{\partial \widetilde{\boldsymbol{u}} \otimes \partial \widetilde{\boldsymbol{u}}}\right|_{\widetilde{\boldsymbol{u}}=\widetilde{\boldsymbol{u}}^n} (\widetilde{\boldsymbol{u}}^{n+1} - \widetilde{\boldsymbol{u}}^n) \approx \boldsymbol{0}, \tag{25}$$

where $\widetilde{\boldsymbol{u}}^n$ is the nodal displacement parameters in the $n-$th iteration step. This equation is the governing equation, and it can be expressed in a more concise form as

$$\boldsymbol{K}(\widetilde{\boldsymbol{u}}^n)(\widetilde{\boldsymbol{u}}^{n+1} - \widetilde{\boldsymbol{u}}^n) = \boldsymbol{P}(\widetilde{\boldsymbol{u}}^n), \tag{26}$$

where

$$\boldsymbol{K}(\widetilde{\boldsymbol{u}}^n) = \left.\frac{\partial^2 U_{\text{int}}(\widetilde{\boldsymbol{u}})}{\partial \widetilde{\boldsymbol{u}} \otimes \partial \widetilde{\boldsymbol{u}}}\right|_{\widetilde{\boldsymbol{u}}=\widetilde{\boldsymbol{u}}^n}$$

and

$$\boldsymbol{P}(\widetilde{\boldsymbol{u}}^n) = -\left.\frac{\partial U_{\text{int}}(\widetilde{\boldsymbol{u}})}{\partial \widetilde{\boldsymbol{u}}}\right|_{\widetilde{\boldsymbol{u}}=\widetilde{\boldsymbol{u}}^n} + \boldsymbol{F} \cdot \boldsymbol{\phi}$$

are the global stiffness matrix and the non-equilibrium force vector, which can be constructed using their corresponding local components $K_0$ and $P_0$, respectively. These local components can be derived by taking the first- and second-order partial derivatives of the cell potential $V_0$ with respect to the displacement variable $\tilde{u}$ as

$$K_0^{\alpha i \beta j} = \frac{\partial^2 V_0}{\partial \tilde{u}_{\alpha i} \partial \tilde{u}_{\beta j}} = \sum_{I=1}^{8} \sum_{J=1}^{8} \frac{\partial^2 V_0}{\partial u_{Ii}^h \partial u_{Jj}^h} \delta_{I\alpha} \delta_{J\beta} \phi_{I\alpha} \phi_{J\beta}, \tag{27}$$

$$P_0^{\alpha i} = -\frac{\partial V_0}{\partial \tilde{u}_{\alpha i}} = -\sum_{I=1}^{8} \frac{\partial V_0}{\partial u_{Ii}^h} \delta_{I\alpha} \phi_{I\alpha}, \tag{28}$$

where $\alpha$ and $\beta$ represent the global node numbers. The $i$ and $j$, which take values from 1 to 3, denote the component indices. The parameter $\delta_{I\alpha}$ is set to 1 if the $\alpha - th$ node lies within the influence domain of the $I - th$ atom; otherwise, it is equal to 0. $\phi_{I\alpha}$ denotes the shape function associated with the $\alpha - th$ node within the influence domain of the $I - th$ atom.

The governing equation presented in Eq. (26) offers a viable avenue for iterative solution through the application of the Newton-Raphson method. A series of extensive numerical tests have been conducted, and the results clearly demonstrate that approximately a dozen iteration steps are sufficient to achieve the minimal energy state for a load step. This finding is of significant importance as it reveals that the governing equation operates at an order of *N*, where *N* represents the number of nodes. In essence, the MAN method can be classified as an order-*N* approach. Compared to order-*N²* numerical methods, such as the conjugate gradient method which is widely employed in atomistic simulations, the MAN method exhibits superior efficiency. Additionally, since nodes can be arranged freely, with the selection of a fewer number of nodes, the efficiency can be further improved.

However, in the context of large-deformation problems, material or geometric nonlinearities often pose a formidable challenge. These nonlinearities can trigger instability in numerical computations. Concurrently, the stiffness matrix $\boldsymbol{K}$ may lose its positive definiteness and become non-positive. When this occurs, the iterative process may fail to converge to the optimal solution, leading to inaccurate or even meaningless results. This troublesome issue can be addressed using stiffness matrix modification methods, such as the '$\boldsymbol{K} + \alpha \boldsymbol{I}$' approach.[41]

## 3. Feasibility of MAN method

To verify the feasibility and effectiveness of the proposed MAN method for investigating the large-deformation and fracture behaviors of 2D CNs under tensile conditions, bilayer graphene and diamane with a length of 28.9 Å (along the armchair direction) and a width of 27.5 Å (along the zigzag direction) are selected as samples. Research reveals that AA-stacked diamane and AB-stacked diamane exhibit nearly identical Young's modulus, shear modulus, Poisson's ratio, and fracture strength.[12, 13] In this study, AA-stacked diamane is chosen as the research sample. To carry out accurate comparative research, this paper adopts the atomic-scale finite element method to perform fully atomistic simulation for comparison.[41] This method enables the selection of the same potential function and loading conditions as those employed in this study, and provides computational accuracy comparable to that of molecular dynamics simulations.[42-45]

In all the following examples, the quasi-static condition is assumed. In numerical simulations, $2(n \times m)$ nodes signify that n and m nodes are respectively distributed along the armchair and zigzag directions of one single continuous sheet layer of a 2D CNs. Prior to loading, the proposed "bilayer graphene + interlayer bonding" model (refer to Fig. 1(a)) is relaxed to achieve an undeformed equilibrium configuration. Subsequently, rigid body translations are applied to the two ends of the multiscale model during the loading process (see Fig. 3 for reference). In each loading step, the two ends of the model first move to new positions, and then the body is driven to its equilibrium configuration by minimizing the potential energy of the atomic system.

Fig. 4 illustrates the variation of the fracture strain of bilayer graphene and diamane with respect to the scaling parameter s in Eq. (13). The loading is applied along the zigzag direction, and $2(23 \times 14)$ nodes are employed, with their number equal to that of the carbon atoms in the system. As can be seen from the figure, for bilayer graphene, the value of s can range from 2.0 to 4.0, while for diamane, the effective range is only between 2.0 and 3.0. This may be due to the complex structure of diamane. When the range of the influence domain expands, the number of nodes within the influence domain increases exponentially, resulting in slow convergence. The cumulative numerical calculation errors arising from this lead to a decline in the accuracy of numerical simulations. In the following research, the value of s will be selected within the range of 2.0-3.0.

In the context of our research on the mechanical properties of bilayer graphene and diamane under zigzag direction uniaxial tension, Fig. 5 illustrates how the number of nodes influences the relationship between strain energy and strain. As shown in Fig.

5(a), when the number and distribution of nodes are consistent with those of atoms, the computational results are in exact accordance with those obtained from fully atomistic simulations. However, as the number of nodes decreases, the fracture strain exhibits an increasing trend. This may be attributed to the fact that, under the same scaling parameter s, the range of the influence domain expands. The deformation of bilayer graphene is averaged over a larger area, thereby smoothing out the local real details. The phenomenon where the local strain energy increases while the strain energy in other regions is released is replaced by a uniform increase in strain energy. Consequently, there is a "false" elevation in the predicted value of the fracture strain. This reflects the characteristic of continuum models, namely, the collective description of the behavior of multiple atoms.

Unlike the computational results for graphene, where a concurrent decline in strain energy and fracture strain of diamane is observed as the number of nodes decreases (as illustrated in Fig. 5(b)). The primary cause of this distinct behavior is that the reduction in nodes leads to distortion in the prediction of diamane's configuration (as shown in Fig. 6). When the number and distribution of nodes are comparable to or exceed those of atoms, the obtained results are in exact accordance with those from fully atomistic simulations. As the number of nodes continues to decrease, an increasing tendency similar to that in graphene emerges.

From the above computational examples, it can be observed that the proposed MAN method is a multi-scale approach. When the number and distribution of nodes are comparable to those of atoms, results consistent with those from atomic methods can be obtained. Conversely, when the number of nodes is less than that of atoms, the method exhibits characteristics of a continuum. Therefore, by adjusting the node arrangement, a smooth cross-scale transition can be readily achieved, thereby avoiding the spurious forces at the interfaces that are common in traditional multi-scale coupling methods.[46-48]

The mechanical responses of diamane under uniaxial tension along the zigzag and armchair directions are depicted in Fig. 7. It can be observed that the fracture strain of diamane is approximately 25.8% along the zigzag direction and about 14.8% along the armchair direction, showing a difference of roughly 50%. This result is in perfect agreement with the findings reported by Wu et al..[13]

It has been reported that the mechanical properties of diamane, such as Young's modulus, shear modulus and Poisson's ratio, are similar to those of diamond.[12] The Young's modulus $E$, shear modulus $G$, and Poisson's ratio $v$ of diamane can be calculated using the following formulas, respectively, as

$$E = \frac{2U_{\text{int}}l}{A\Delta l^2}, \qquad (29)$$

$$G = \frac{E}{2(1+v)}, \qquad (30)$$

$$v = \frac{\varepsilon_x}{\varepsilon_y}, \qquad (31)$$

where $A$ represents the cross-sectional area, with the thickness of diamane taken as 4.6 Å.[12] $\varepsilon_x$ denotes the strain in the loading direction, while $\varepsilon_y$ represents the strain in the perpendicular direction. Table 1 presents the Young's modulus, shear modulus, and Poisson's ratio of diamane along both the zigzag and armchair directions. During the initial elastic stage, the values of Young's modulus, shear modulus, and Poisson's ratio along the zigzag direction are 1285.11 GPa, 608.48 GPa, and 0.056, respectively, while those along the armchair direction are 1271.52 GPa, 598.08 GPa, and 0.063, respectively. These values exhibit excellent consistency with the findings reported in the literature.[12] The Young's modulus and shear modulus are nearly identical along both the zigzag and armchair directions and demonstrate a declining trend with increasing loading, which reveals the material nonlinearity characteristics of diamane.

Therefore, it can be concluded that the proposed MAN method can accurately predict the nonlinear large-deformation and fracture behaviors of bilayer graphene composed of $sp^2$ bonds and diamane composed of $sp^3$ bonds.

## 4. Effect of interlayer bonds on fracture behavior of 2D CNs

The aforementioned research findings reveal significant differences in the tensile fracture mechanical properties between bilayer graphene and diamane, as illustrated in Fig. 5. From a structural perspective, the primary distinction between the two lies in the presence of interlayer $sp^3$ bonds. Despite being oriented perpendicular to the loading direction, the introduction of these interlayer bonds modifies the bond lengths and bond angles of the in-plane bonds, thereby influencing their tensile mechanical properties. In this section, we continue to focus on 2D CNs of the same scale as those studied in Section 3, exploring the impact of varying the number of interlayer bonds on the tensile fracture properties of 2D CNs.

First, we investigate the influence of $sp^3$ interlayer bonds on loading in the zigzag direction. The arrangements of six different patterns of interlayer $sp^3$ bonds are shown in Fig. 8. The fracture strains of Mod-P 1-6 are 20.3%, 20.2%, 20.2%, 20.2%, 20.1%, and 20.1%, respectively. It can be seen that an increase in the number of interlayer $sp^3$ bonds does not enhance the structural resistance to damage. The fracture configurations are illustrated in Fig. 9, where cracks initiate at the corner points and propagate through the $sp^3$ bonds. The fracture strains of Mod-P 1-6 are all around 20%, slightly lower than the fracture strain of bilayer graphene, which is 21.8% (see Fig. 4). This may be attributed to the irregular warping at the edges of graphene caused by the introduction of $sp^3$ bonds (see Fig. 8), thereby reducing its resistance to damage.

The arrangements of six different linearly aligned interlayer $sp^3$ bonds are presented in Fig. 10. The interlayer $sp^3$ bonds in Mod-L 1-3 are arranged parallel to the loading direction, with fracture strains of 20.1%, 20.2%, and 20.1%, respectively, and their fracture configurations are shown in Fig. 11(a)-(c). It is evident that variations in the alignment position have negligible effects on the fracture strain of the structure. The interlayer $sp^3$ bonds in Mod-L 4-6 are arranged perpendicular to the loading direction, with fracture strains of 17%, 17.1%, and 17%, respectively. These values are significantly lower than the approximately 20% for Mod-L 1-3 and the 21.8% for bilayer graphene. The fracture configurations are depicted in Fig. 11(d)-(f), where fracture initiation does not occur at the corner points but rather at the $sp^3$ bonds near the boundary. The crack surfaces propagate through all the $sp^3$ bonds in a highly regular manner. A possible explanation is that the combination of $sp^3$ bonds near the boundary and the zigzag boundary forms pre-embedded angular "notches," which readily facilitate the formation of tear-type cracks during loading, thereby reducing the structural resistance to damage.

Next, full rows of sp³ bonds are arranged along the zigzag direction, with the row count increasing incrementally from the center to both ends. The corresponding equilibrium configurations of 2D CNs are shown in Fig. 12. The impact of varying numbers of sp³-bond rows on the fracture behavior of 2D CNs is illustrated in Fig. 13. As depicted in the figure, when two rows of sp³ bonds are inserted at the center, the tensile strength of the structure is actually lower than that of the zero-row case (bilayer graphene). As mentioned earlier, this is possibly because the incorporation of two rows of sp³ bonds induces irregular warping of the graphene (as shown in Fig. 12). The combined effects of boundary warping and necking cause localized stretching and subsequent fracture of the sp² bonds near the corner points. As the number of interlayer-bond rows increases and warping simultaneously weakens, the tensile strength of 2D CNs gradually approaches that of diamane (with 14 rows of sp³ bonds).

The crack propagation behavior of the 2D CNs is illustrated in Fig. 14. Initially, the crack originates from the corner point at the end of the lower layer and then propagates along the zigzag direction, as shown in Fig. 14(a). Upon encountering the interlayer bonds, the crack transfers from the lower layer to the upper layer, causing the upper layer to rupture from the middle, as depicted in Fig. 14(b). Subsequently, the crack continues to propagate separately in both the upper and lower layers, as presented in Fig. 14(c) and (d), until the structure ultimately fails.

The above research findings reveal that edge warping adversely affects the tensile strength of 2D CNs. Would introducing interlayer bonds at the boundaries of 2D CNs to mitigate edge warping enhance its tensile properties? To further investigate this issue, four rows of sp³ interlayer bonds were arranged in two distinct configurations within the 2D CNs: one with bonds concentrated at both ends, and the other with bonds evenly dispersed throughout (see Fig. 15). Through calculations, it has been found that arranging interlayer bonds at the boundaries can effectively suppress boundary warping. Although the strength of sp³ hybridization is higher than that of sp² hybridization, the fracture strain remains at 20%, which is still lower than that of bilayer graphene (21.8%). A possible reason is that the boundary necking effect plays a dominant role, causing non-uniform deformation of sp³ bonds. Specifically, the bond lengths near the boundary corners are longer, while those in the middle are shorter, making the atomic bonds at both ends more prone to fracture and leading to crack formation. In contrast, for the pattern with uniformly distributed bonds, although there is still some warping at the boundaries, the sp² bonds at the boundaries are reinforced by adjacent sp³ bonds, resulting in more uniform deformation. This, to a certain extent, counteracts the non-uniform deformation caused by the necking effect, enabling the fracture strain to reach 22.9%. It can thus be seen that to improve the tensile strength of 2D CNs, both the reduction of boundary warping and the mitigation of necking effect need to be considered simultaneously.

Next, we will investigate the impact of sp³ interlayer bond arrangements on the 2D CNs under tension along the armchair direction. Fig. 16 illustrates the influence of the number of rows of sp³ interlayer bonds along the armchair direction on the fracture strain of 2D CNs when subjected to tension in the armchair direction. The fracture strains of bilayer graphene and diamane under loading in the armchair direction are 14.1% and 14.7%, respectively, with only a slight difference between them. Therefore, the addition of interlayer bonds does not significantly improve the tensile properties of 2D CNs.

The corresponding crack propagation configurations are depicted in Fig. 17. The crack initiates and propagates along the zigzag direction, as illustrated in Fig. 17(a). As the crack advances and encounters the interlayer bonds, it undergoes a transfer process. Just as in the zigzag direction loading, the crack shifts from one layer to the adjacent layer. This transfer causes the receiving layer to rupture, starting from a central region, as shown in Fig. 17(b). After this interlayer transfer and rupture event, the crack continues to propagate independently within both layers. This stage of independent crack propagation in the two layers is presented in Fig. 17(c) and (d). The crack keeps extending in each layer until the entire structure eventually fails. Notably, the crack also propagates along the zigzag direction, consistent with the scenario under zigzag loading, indicating no correlation with the loading direction. Interestingly, the crack does not initiate at the boundary but originates from the interior adjacent to the boundary. The underlying reason is that the necking effect at the boundary induces the movement and rotation of specific boundary bonds b, d, and f (refer to Fig. 18). This, in turn, reduces the probability of other boundary bonds a, c, and e undergoing fracture. Consequently, bond g, which is located farther from the boundary, becomes the most susceptible to fracture.

## 4. Concluding remarks

In order to investigate the effect of interlayer $sp^3$ bonds on the nonlinear large-deformation and fracture behaviors of 2D CNs under tensile conditions, this paper develops a MAN method. This method can achieve a smooth transition from the atomic scale to the continuum scale by adjusting the node arrangement, thereby avoiding the common problem of interfacial spurious forces in traditional multiscale coupling methods. Based on this method, it is possible to accurately calculate the material parameters of 2D CNs and predict their large-deformation behavior as well as fracture behavior under tension.

Our research reveals that the quantity and distribution of interlayer $sp^3$ bonds have a significant impact on the fracture behavior of 2D CNs. Despite the fact that $sp^3$ hybridization exhibits greater strength compared to $sp^2$ hybridization, the incorporation of $sp^3$ bonds do not guarantee an improvement in the tensile performance of 2D CNs. It is also necessary to take into account the influence of boundary warping and necking on the tensile strength of these nanostructures. In contrast, a uniformly distributed pattern of $sp^3$ bonds proves beneficial in mitigating strain localization, thereby enhancing the overall structural strength of 2D CNs.

**Acknowledgments**

This work was supported by the National Natural Science Foundation of China (12172158); and the Shandong Provincial Natural Science Foundation, China (ZR2024QE054) .

**Figures and caption**

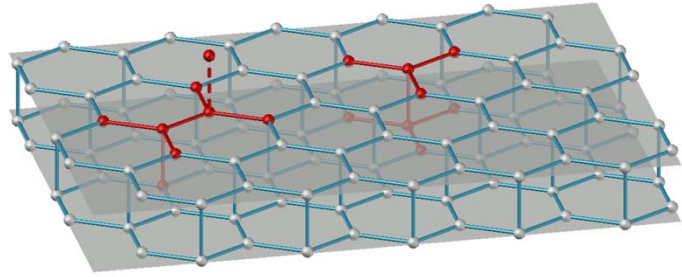

(a)

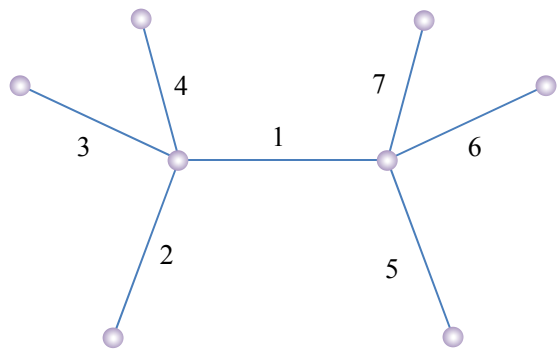

(b)

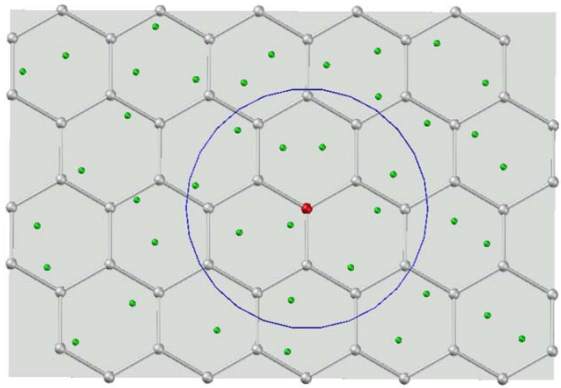

(c)

Fig. 1. A schematic diagram of MAN model: (a) the "bilayer graphene + interlayer bonding" model, where the red bonds indicate sp² and sp³ hybridization, and the gray sheets represent virtual continuum sheets, (b) the sp³ hybridized representative cell, where the numbers indicate the bond numbers, and (c) the influence domain of a carbon atom, where the red point denotes an atom and the green points denote nodes.

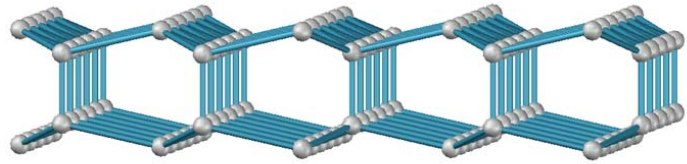

Fig. 2. Schematic illustration of the undeformed equilibrium configuration of AA-stacked diamane.

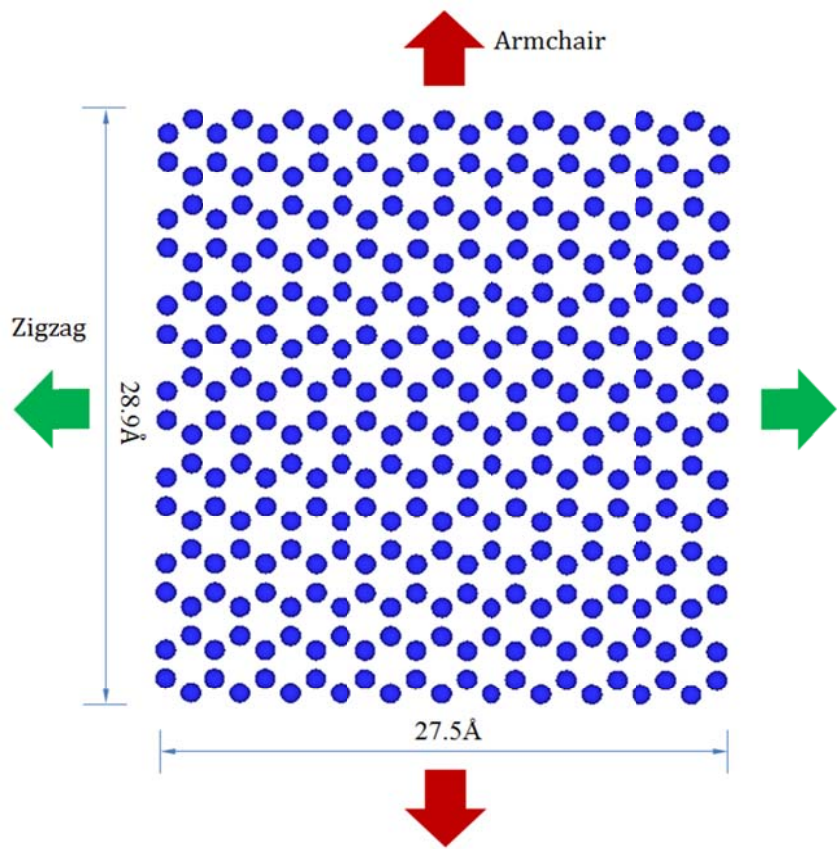

Fig. 3. Schematic diagram of uniaxial stretching of the bilayer graphene/diamane along the armchair/zigzag direction, respectively.

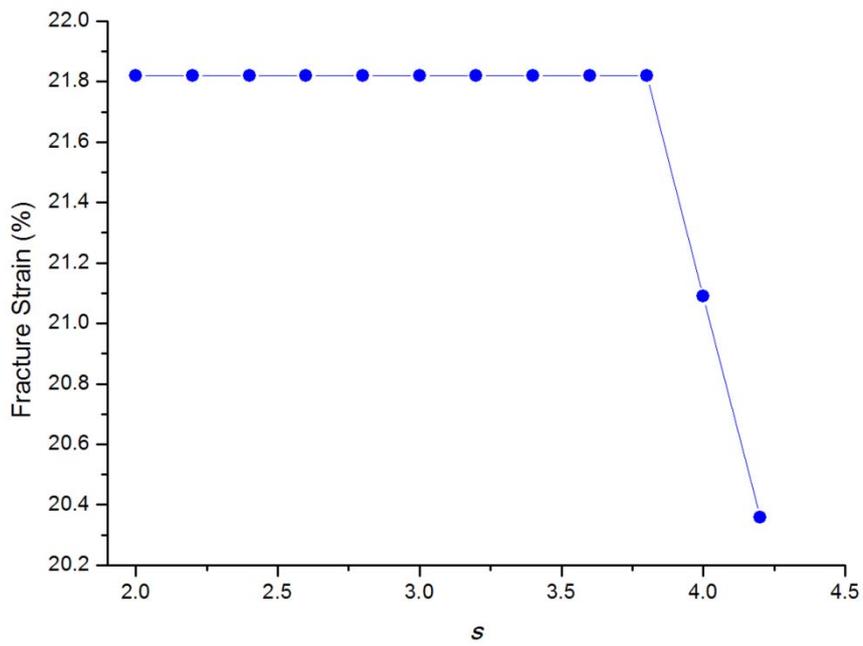

(a)

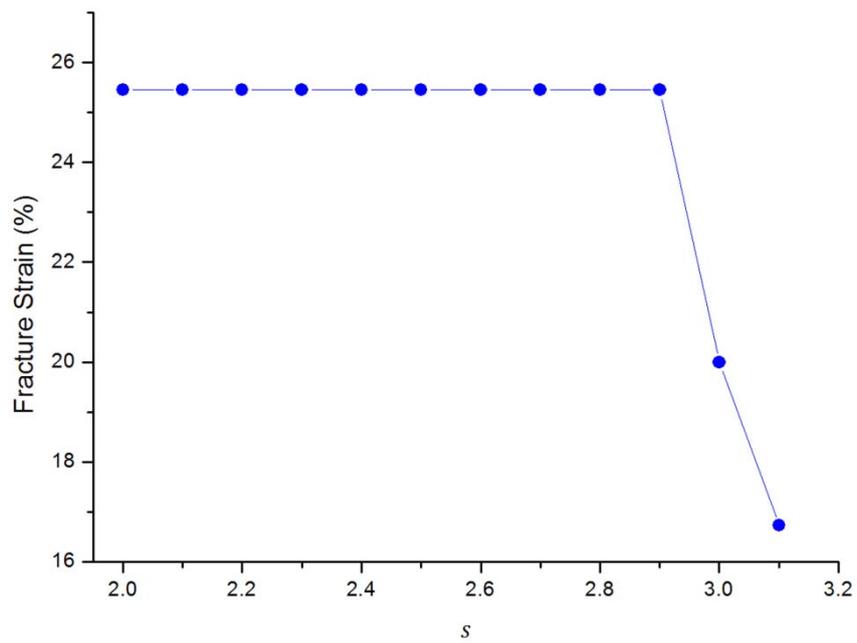

(b)

Fig. 4. The variation of the fracture strain of (a) bilayer graphene and (b) diamane with respect to the scaling parameter $s$.

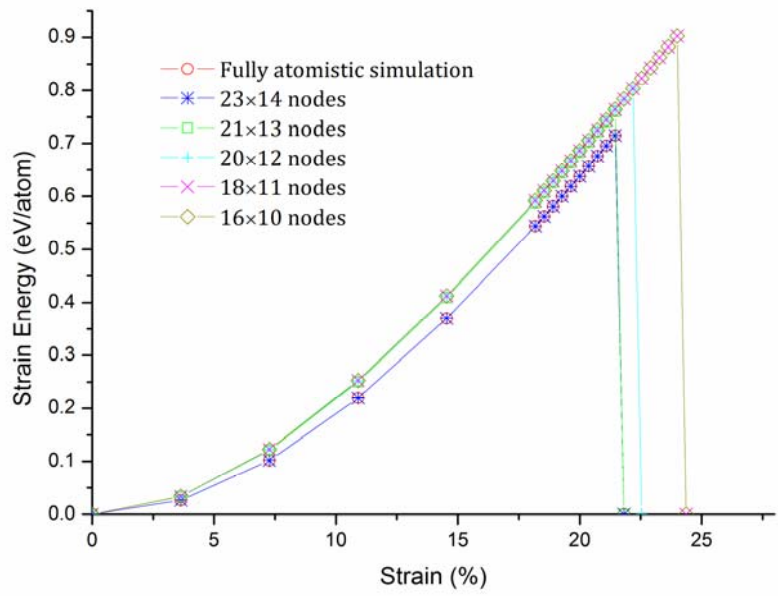

(a)

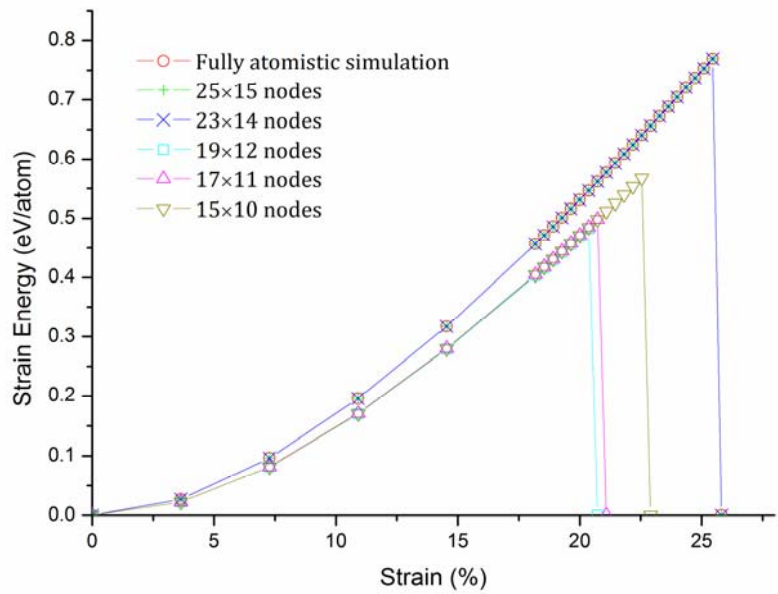

(b)

Fig. 5. Influence of the number of nodes on the relationship between strain energy and strain for the (a) bilayer graphene and (b) diamane under zigzag direction uniaxial tension.

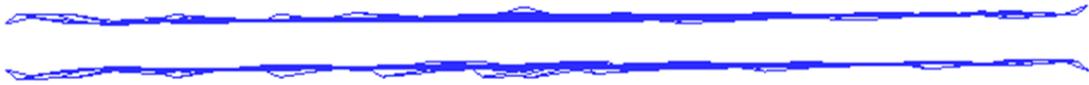

(a)

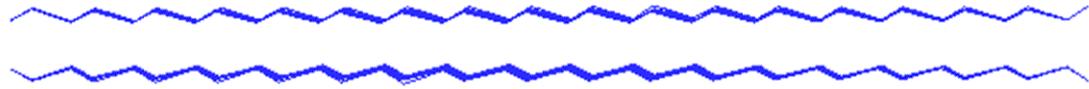

(b)

Fig. 6. Lateral view of the predicted structural configuration of the diamane with (a) 2(17×11) nodes and (b) 2(23×14) nodes.

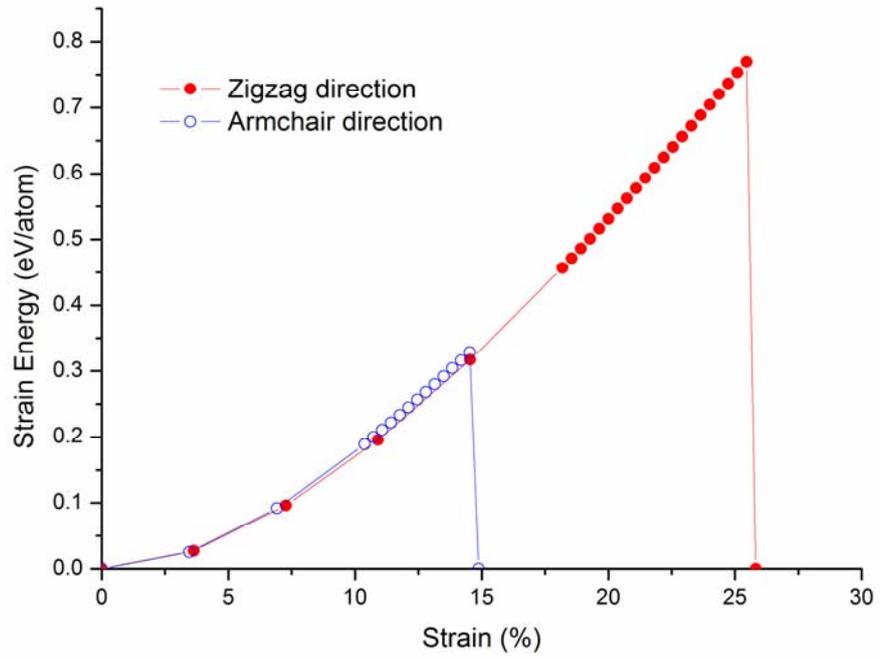

Fig. 7. The relationship between the strain energy and strain of diamane under different loading directions.

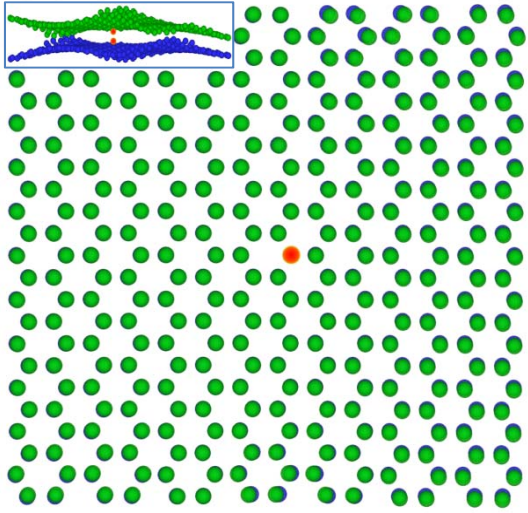
(a)

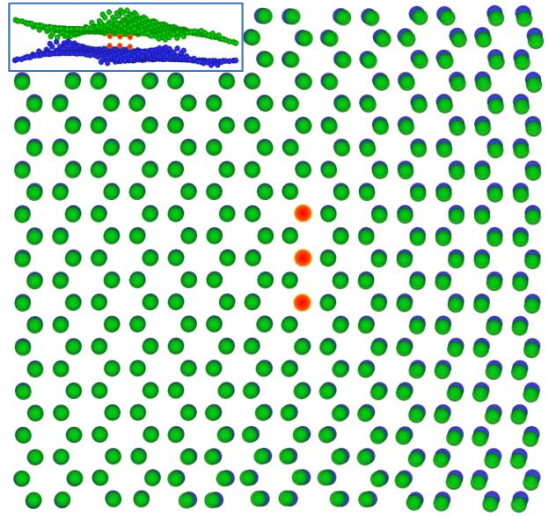
(b)

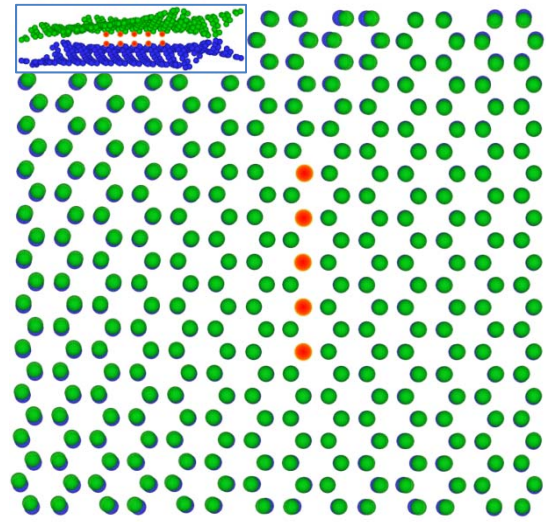
(c)

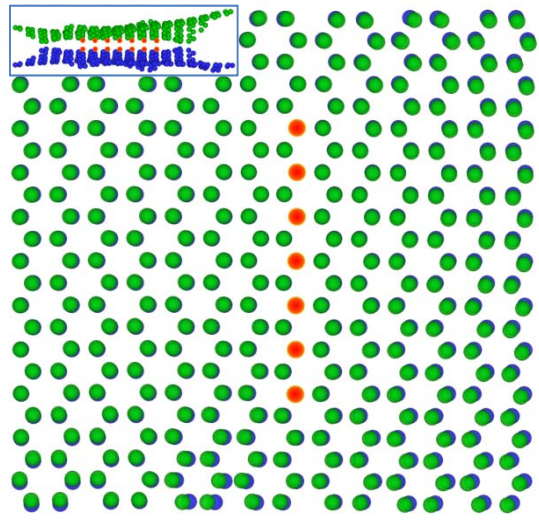
(d)

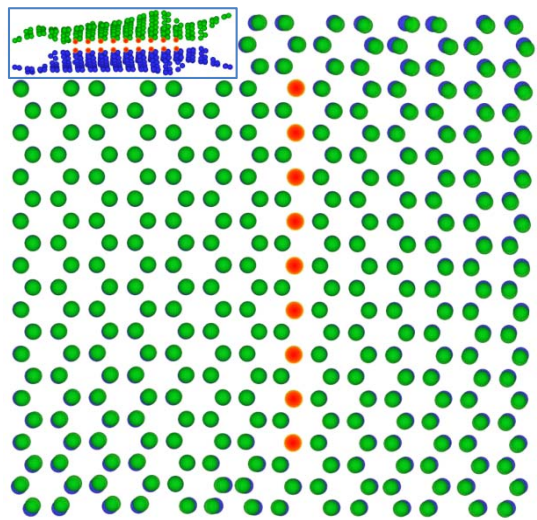
(e)

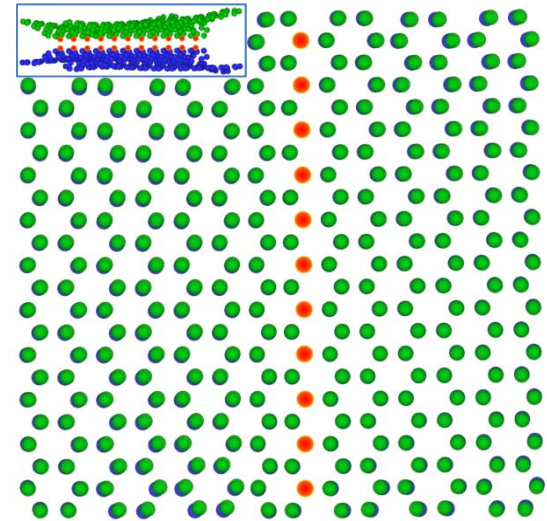
(f)

Fig. 8. The initial configurations of different patterns of interlayer $sp^3$ bonds: (a) Mod-P 1, (b) Mod-P 2, (c) Mod-P 3, (d) Mod-P 4, (e) Mod-P 5, and (f) Mod-P 6. In these configurations, red points represent the atoms that are connected by interlayer sp³ bonds.

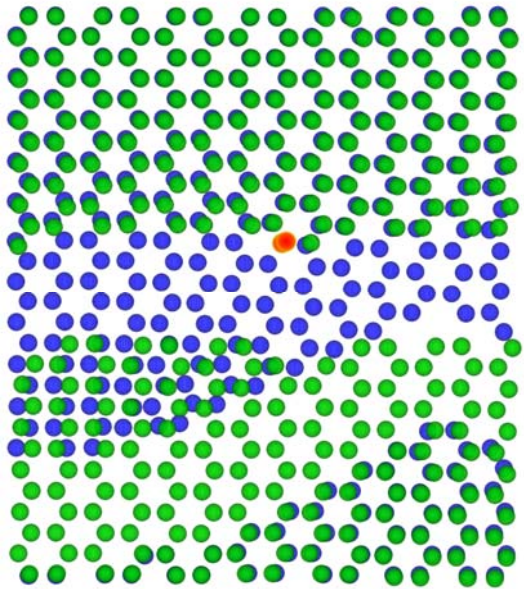
(a)

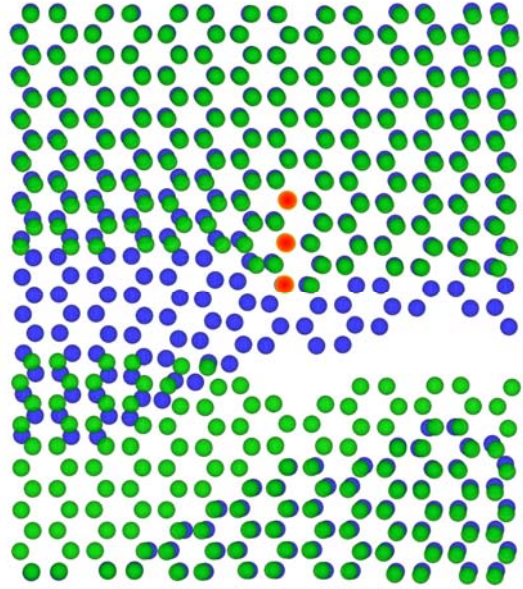
(b)

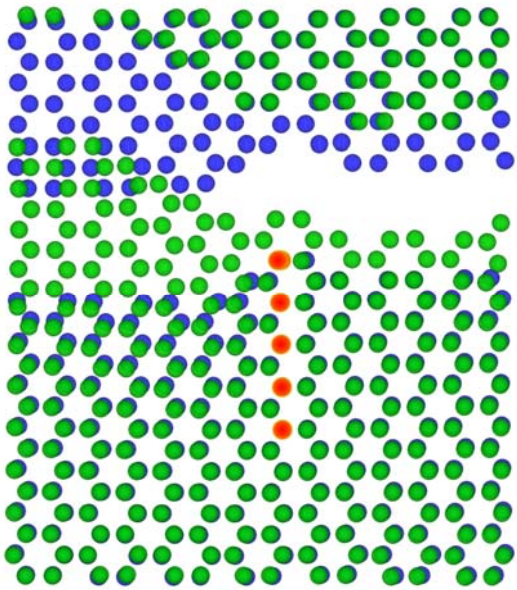
(c)

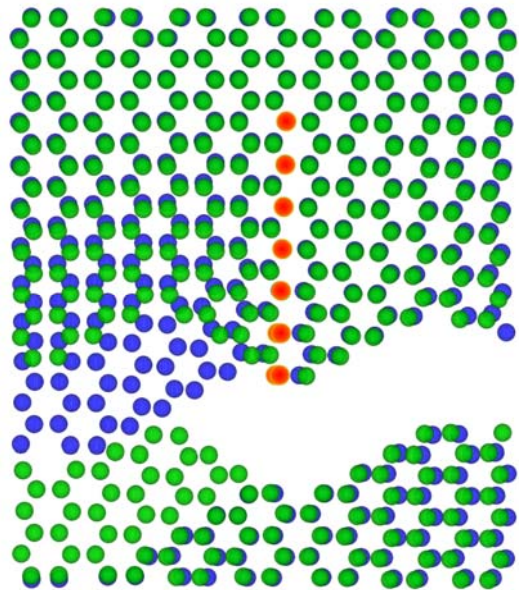
(d)

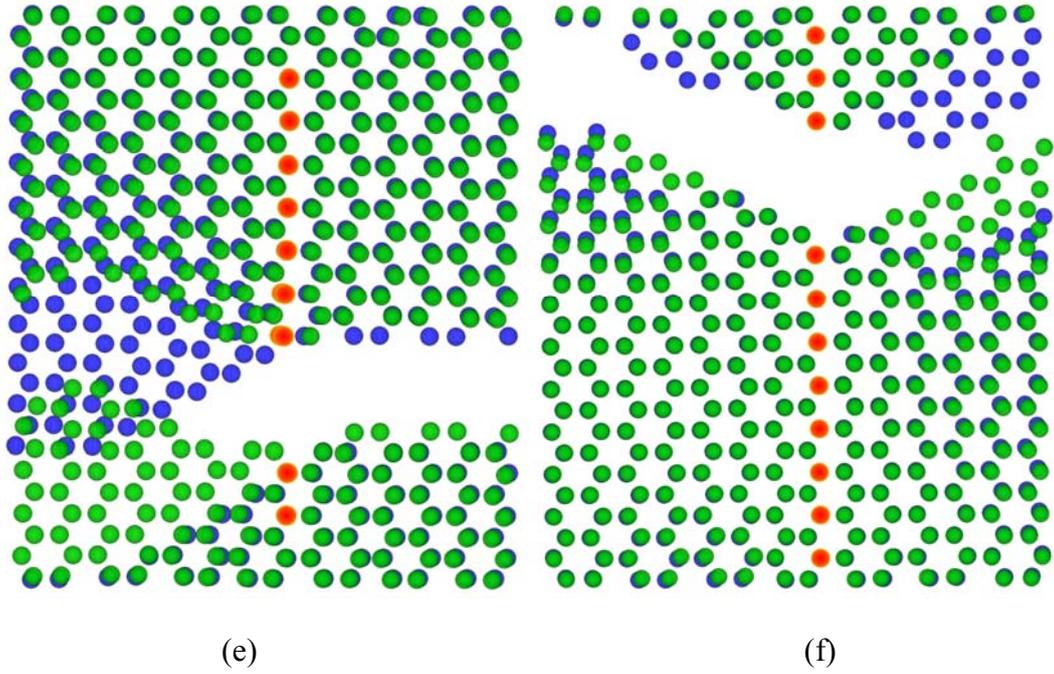

(e) (f)

Fig. 9. The fracture configurations of (a) Mod-P 1, (b) Mod-P 2, (c) Mod-P 3, (d) Mod-P 4, (e) Mod-P 5, and (f) Mod-P 6, respectively.

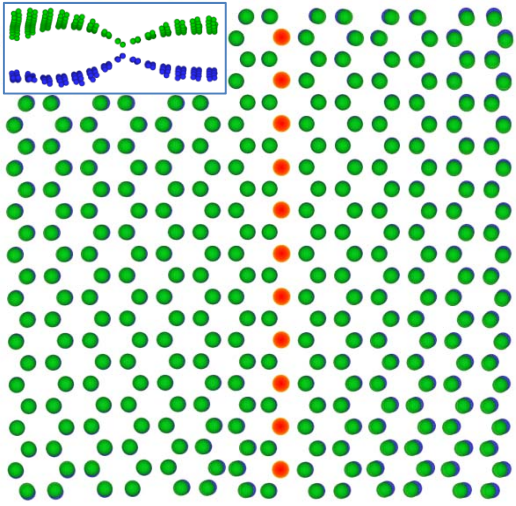
(a)

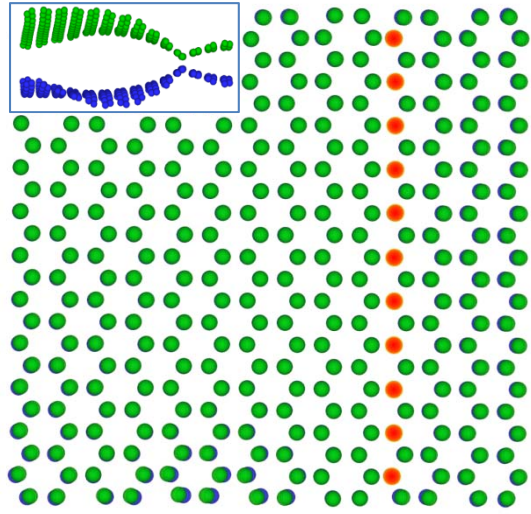
(b)

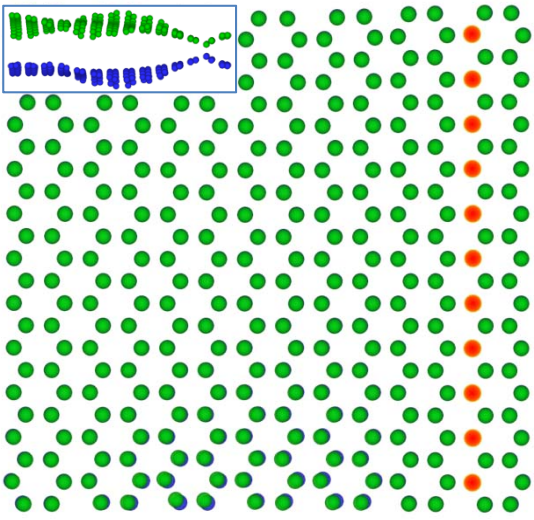
(c)

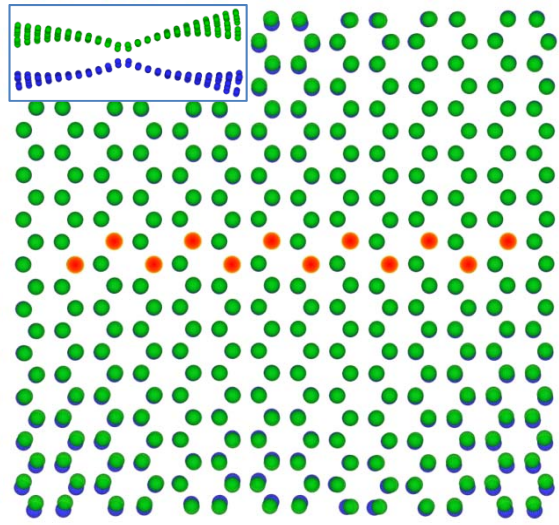
(d)

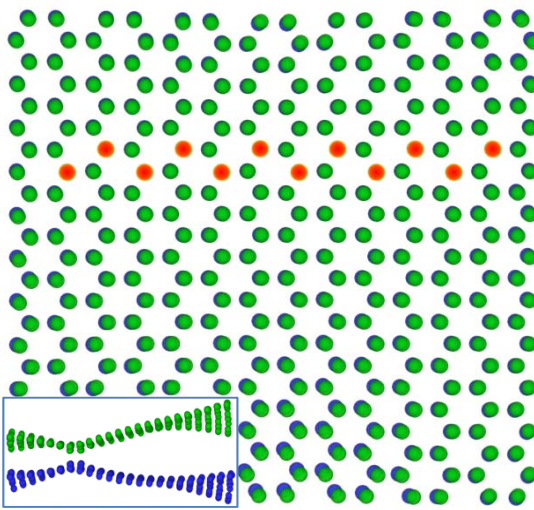
(e)

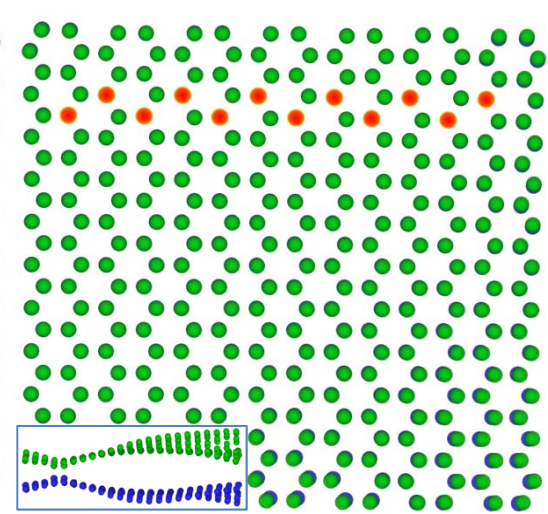
(f)

Fig. 10. The initial configurations for various linear arrangements of interlayer $sp^3$ bonds: (a) Mod-L 1, (b) Mod-L 2, (c) Mod-L 3, (d) Mod-L 4, (e) Mod-L 5, and (f) Mod-L 6. The red points in these configurations indicate the atoms that are interconnected through interlayer $sp^3$ bonds.

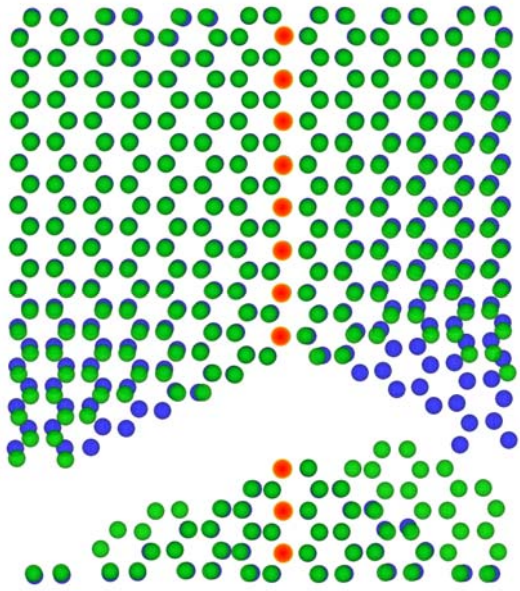 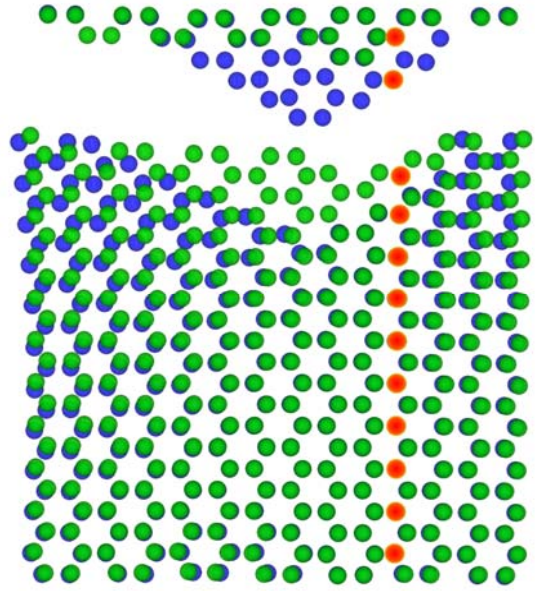

(a) (b)

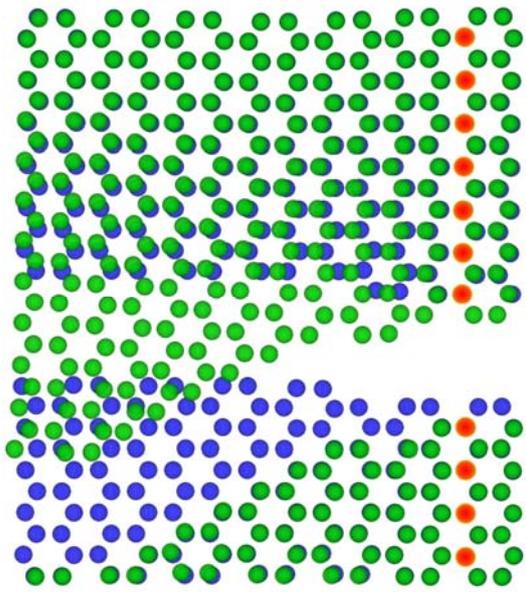 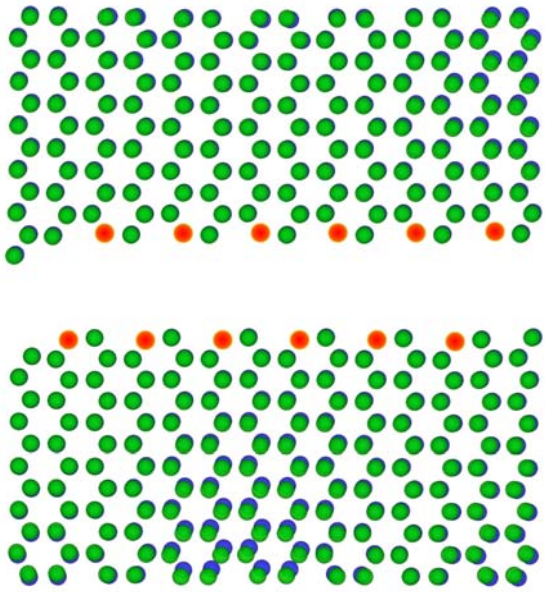

(c) (d)

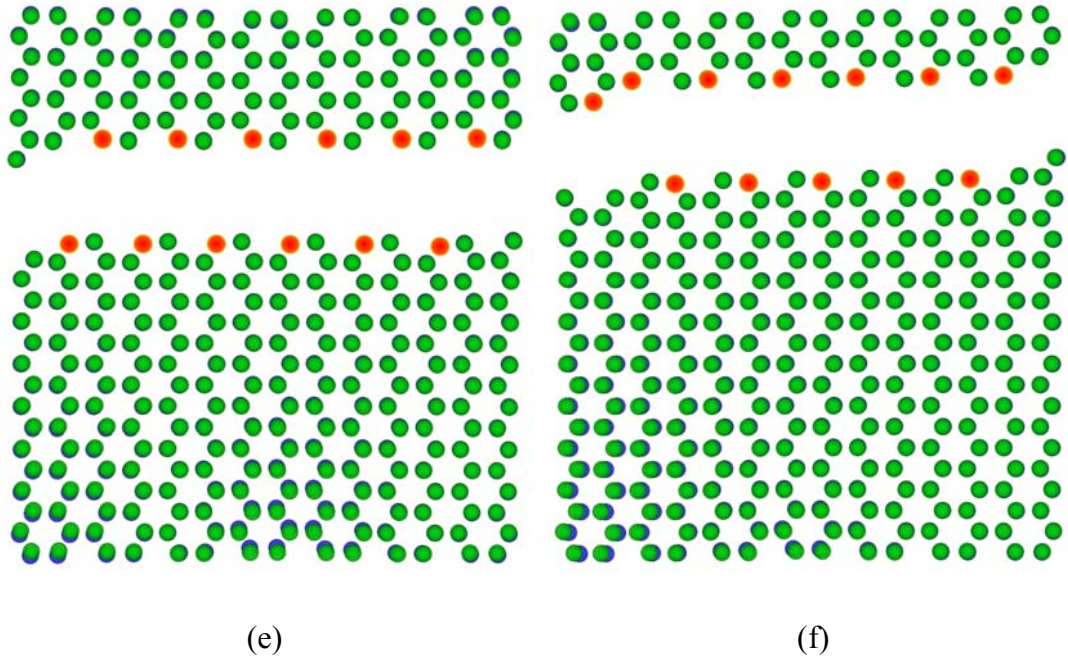

(e)            (f)

Fig. 11. The fracture configurations of (a) Mod-L 1, (b) Mod-L 2, (c) Mod-L 3, (d) Mod-L 4, (e) Mod-L 5, and (f) Mod-L 6, respectively.

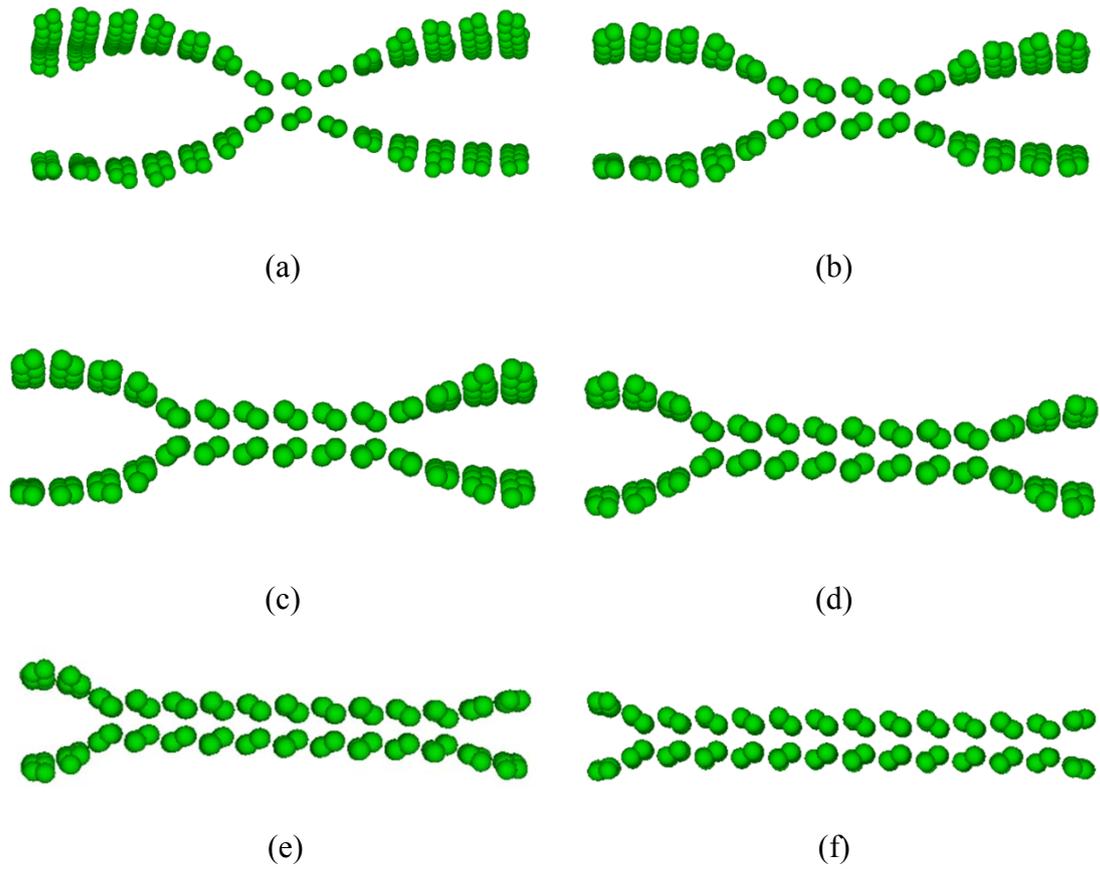

Fig. 12. Side views of the initial equilibrium configurations of the 2D CNs with different numbers of rows of sp³ interlayer bonds: (a) 2 rows, (b) 4 rows, (c) 6 rows, (d) 8 rows, (e) 10 rows, and (f) 12 rows.

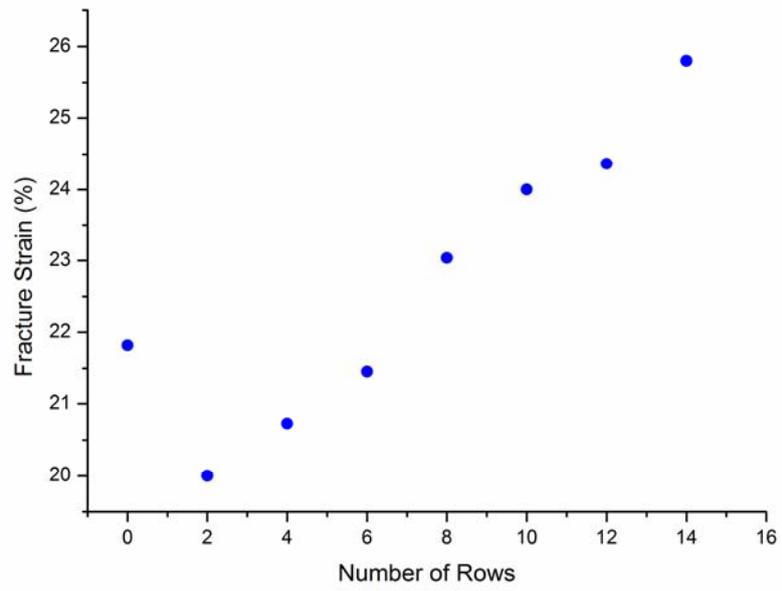

Fig. 13. The influence of number of rows of sp³ interlayer bonds on the fracture strain of 2D CNs under zigzag direction tension.

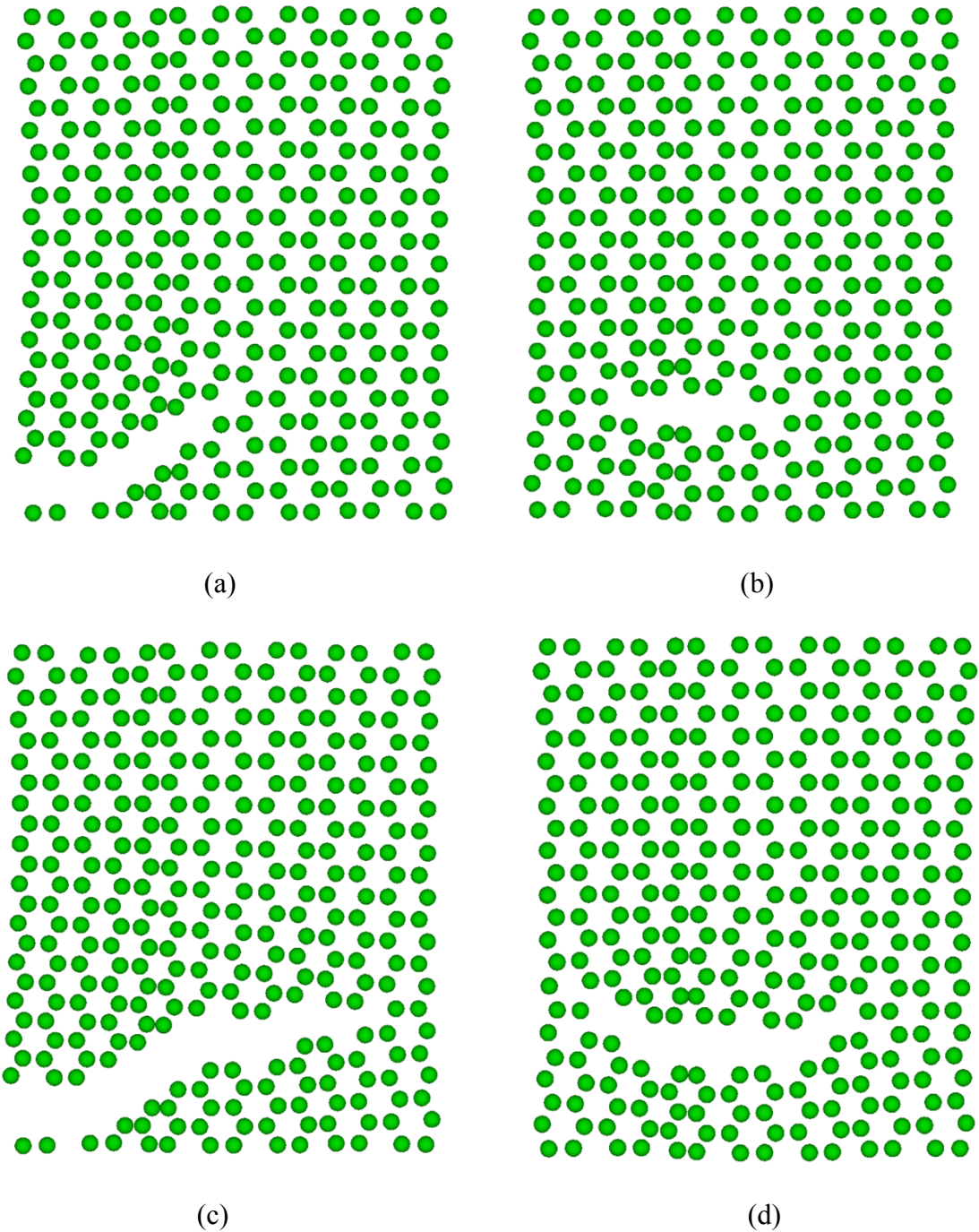

Fig. 14. Crack propagation configurations of the 2D CNs with two rows of sp$^3$ interlayer bonds: (a) and (b) represent the configurations of the lower and upper graphene layers, respectively, at the initial stage of crack propagation; (c) and (d) represent the configurations of the lower and upper graphene layers, respectively, at the advanced stage of crack propagation.

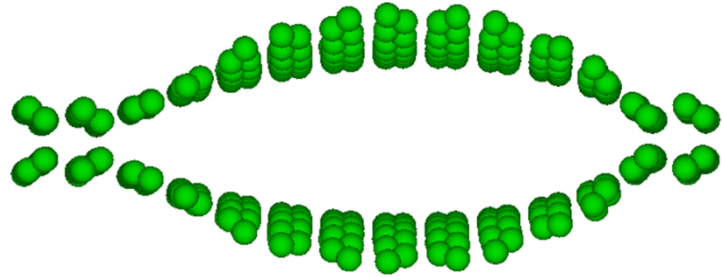

(a)

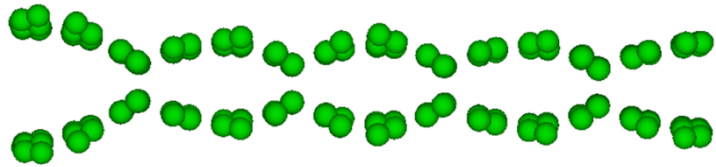

(b)

Fig. 15. Side views of the initial equilibrium configurations of the 2D CNs with 4 rows of sp³ interlayer bonds, showing (a) end-arranged configuration and (b) dispersed-arranged configuration.

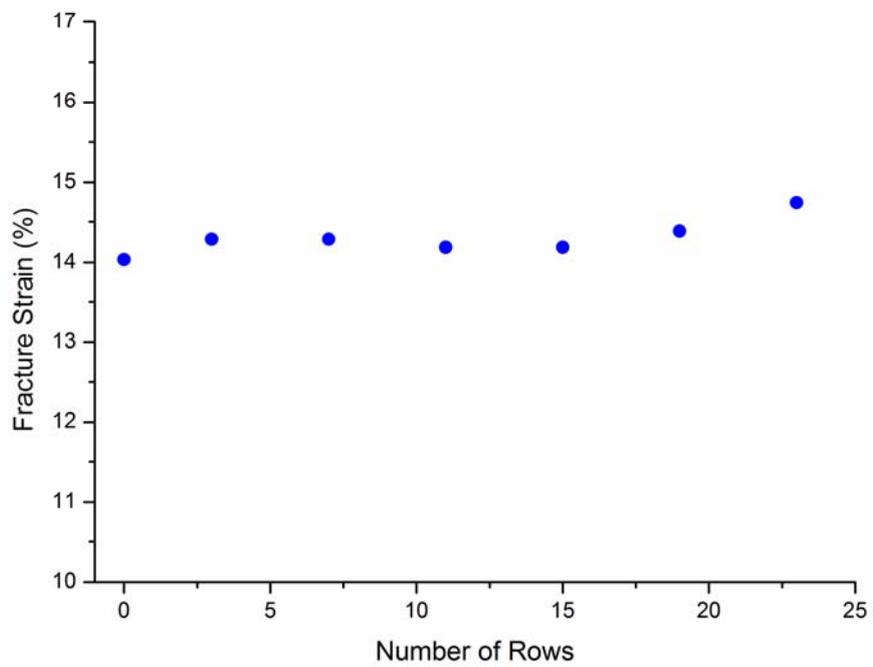

Fig. 16. The influence of number of rows of sp³ interlayer bonds on the fracture strain of 2D CNs under armchair direction tension.

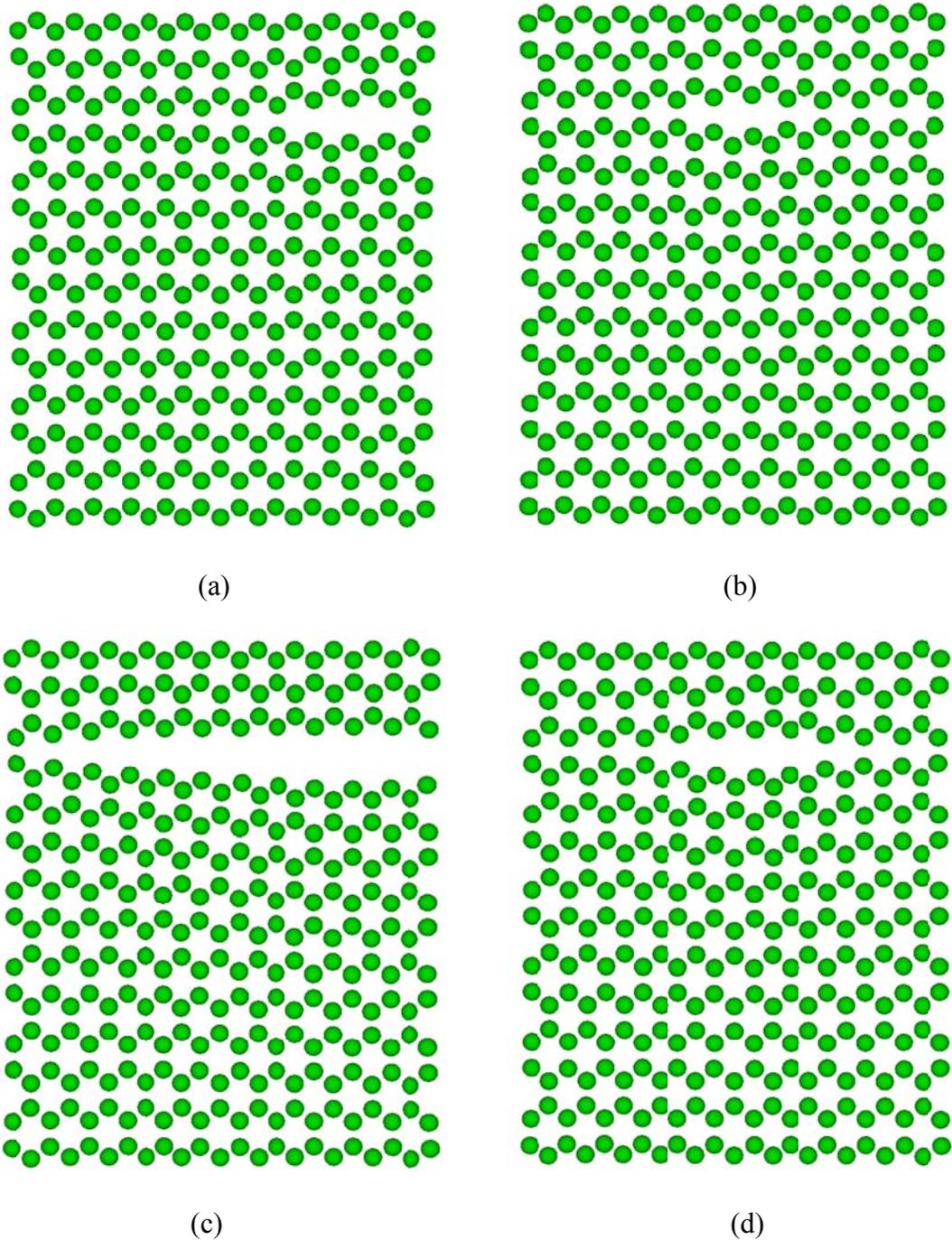

Fig. 17. Crack propagation configurations of the 2D CNs under zigzag direction tension: (a) and (b) represent the configurations of the lower and upper graphene layers, respectively, at the initial stage of crack propagation; (c) and (d) represent the configurations of the lower and upper graphene layers, respectively, at the advanced stage of crack propagation.

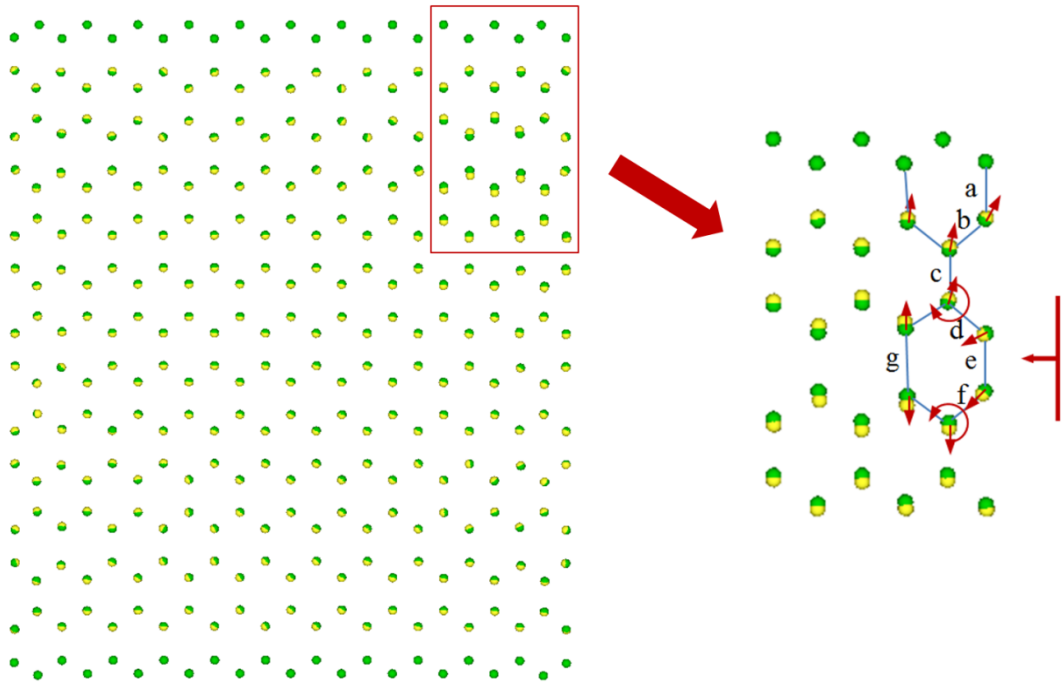

Fig. 18. The trend of crack initiation, where green dots indicate the deformed configuration from the previous step, and yellow dots denote the current-step configuration.

Table 1. The values of Young's modulus $E$, shear modulus $G$ and Poisson's ratio $\nu$ of the diamane.

| | Zigzag direction | | | | Armchair direction | | |
|---|---|---|---|---|---|---|---|
| $\varepsilon$ | $E$ (GPa) | $G$ (GPa) | $\nu$ | $\varepsilon$ | $E$ (GPa) | $G$ (GPa) | $\nu$ |
| $3.64 \times 10^{-3}$ | 1285.11 | 608.48 | 0.056 | $3.42 \times 10^{-3}$ | 1271.52 | 598.08 | 0.063 |
| $7.29 \times 10^{-3}$ | 1259.09 | 595.60 | 0.057 | $6.85 \times 10^{-3}$ | 1270.01 | 599.06 | 0.06 |
| $1.09 \times 10^{-2}$ | 1241.20 | 586.58 | 0.058 | $1.03 \times 10^{-2}$ | 1262.16 | 596.48 | 0.058 |
| $1.46 \times 10^{-2}$ | 1225.48 | 578.06 | 0.06 | $1.37 \times 10^{-2}$ | 1252.77 | 593.73 | 0.055 |
| $1.82 \times 10^{-2}$ | 1210.74 | 570.57 | 0.061 | $1.71 \times 10^{-2}$ | 1242.79 | 590.12 | 0.053 |
| $2.19 \times 10^{-2}$ | 1196.57 | 562.83 | 0.063 | $1.05 \times 10^{-2}$ | 1232.54 | 586.92 | 0.05 |